\documentstyle[12pt, epsf]{article}
\begin{document}
\title{
Dynamical Mass and Parity Condensate \\
in Varying Topological Mass  
}
\author{
Toyoki Matsuyama 
\thanks{e-mail address: matsuyat@nara-edu.ac.jp} \\
Department of Physics, Nara University of Education \\
Takabatake-cho, Nara 630-8528, JAPAN \\
and \\
Hideko Nagahiro \\
Department of Physics, Nara Women's University  \\
Nara 630-8506, JAPAN
}

\maketitle

\begin{abstract}
We investigate how a dynamical mass of a fermion is affected by a 
topological mass of a gauge field in a Maxwell-Chern-Simons $QED_3$ coupled 
with a two-component fermion.  
The dynamical mass and also a parity condensate are estimated by using a 
non-perturbative Schwinger-Dyson method.  
In particular, we study a limit of vanishing the topological mass in detail 
and clarify a linking between theories with and without a Chern-Simons 
term in a non-perturbative level.  
\end{abstract}

\section{Introduction}

In a sense, a space-time with dimensions "2+1" is mysterious because the 
mathematics tells us that there exists a specific term called a 
Chern-Simons term \cite{CS}.  
This term is allowed just in (2+1) dimensions (generally in odd space-time 
dimensions).  
As is well known, a Lagrangian density for an electromagnetic field is 
given by a Maxwell term, which is 1) gauge invariant, 2) Lorentz 
invariant, and 3) bilinear for the gauge field.  
The (abelian) Chern-Simons term also satisfies all of 1) $\sim$ 3).  
Therefore the Maxwell theory in the (2+1) dimensions has a natural extension 
which is defined by adding the Chern-Simons term to the Maxwell Lagrangian.  
This extended version is called a Maxwell-Chern-Simons 
theory. \cite{SS,DJT}  

The modern technology of engineering makes possible to produce low-dimensional 
electron systems in realistic electronic devises.  
Especially, in (2+1)-dimensional systems, novel phenomena as the quantum 
Hall effect \cite{QH} and the high-$T_C$ superconductivity \cite{HTC} 
were discovered.  
It is plausible that these phenomena may have their origins in the dimensions 
of the space-time.  

In fact, there appeared many approaches to understand these macroscopic 
quantum effects by using (2+1)-dimensional quantum field theories with and 
without the Chern-Simons term.  
For example, the (2+1)-dimensional quantum electrodynamics ($QED_3$) was 
used to explain the quantum Hall effect and $QED_3$ with the 
Chern-Simons term (Maxwell-Chern-Simons $QED_3$) provided an anyon model 
which was expected to give us an essential mechanism for the high-$T_C$ 
superconductivity. \cite{PGWF}  
These investigations produced important results and are still in progress.  

What is the physical meaning of the Chern-Simons term?  
The Chern-Simons term gives the gauge field a mass without breaking the gauge 
symmetry. \cite{SS,DJT}  
This mass is called a topological mass because the Chern-Simons term has a 
topological meaning as the secondary characteristic class. \cite{CS}  
Whether the gauge field is massless or massive affects the nature of 
interactions, e.g., the range of interactions.  
In this case, the most important effect of the massive gauge field is to 
rescue the (2+1)-dimensional Maxwell theory from the infrared catastrophe 
which appears in a self-energy of fermion when the Maxwell field interacts 
with matters. \cite{JTAPRS}  

On the other hand, it is known that nonperturbative radiative corrections can 
produce a mass of fermion called a dynamical mass.  
The dynamical mass generation of four-component fermions in $QED_3$ 
without the Chern-Simons term has been studied in Ref.\cite{PABCWABKW}.  
One four-component fermion is equivalent to two two-component fermions.  
The mass term of the two-component fermion breaks the parity ($P$) 
symmetry while the one of the four-component fermion breaks $P \times 
Z_2{\rm (flavour)}$ symmetry.  
In Ref.\cite{HMH}, one of the present authors and others have investigated the 
dynamical mass generation of a single two-component fermion in $QED_3$ 
without the Chern-Simons term.  
A parity-breaking solution which generates the dynamical mass has been 
found .   

Both analyses have been extended to the cases with the Chern-Simons term.  
The study of the four-component dynamical mass in the Maxwell-Chern-Simons 
$QED_3$ has been done in Ref.\cite{HPKKKM}.  
The dynamical mass generation of a single two-component fermion in 
the Maxwell-Chern-Simons $QED_3$ has been studied in Ref.\cite{MNU}.  

The study of Ref.\cite{MNU} is motivated to clarify a role of the topological 
mass in the nonperturbative dynamics.  
The dependence of the dynamical mass on the topological mass has been 
investigated.  
However, as was pointed out there, the numerical estimation of 
the dynamical mass for a very small but non-zero value of the topological 
mass is very difficult technically.  
In this paper, we devise a way of estimation and extend the previous 
analysis further to the case in which the topological mass has much 
more smaller value.  
In addition, we evaluate a parity condensate which is an important quantity 
of indicating an amount of the parity breaking.  
We make clear a linking between theories with and without the Chern-Simons 
term in a nonperturbative level.  

This paper is organized as follows.  
In Sec. 2, we explain the Maxwell-Chern-Simons $QED_3$ briefly.  
By examining a fermion self-energy in a perturbation, 
we demonstrate why we need a nonperturbative analysis for our purpose 
in Sec. 3.  
The Schwinger-Dyson equations are derived and studied by an approximated 
analytical method in Sec. 4.  
We solve the equations by a numerical method in Sec. 5, where the dynamical 
mass and a parity condensate are estimated.  
In Sec. 6, we summarize our results.  

\section{Maxwell-Chern-Simons $QED_3$}

We consider the Maxwell-Chern-Simons $QED_3$ with the two-component Dirac 
fermion. \cite{SS,DJT}  
The Lagrangian density of the theory is given by 
\begin{eqnarray}
{\cal L}= - \frac{1}{4} F_{\mu\nu} F^{\mu\nu}
          + \frac{\mu}{2} \varepsilon^{\mu \nu \rho} A_\mu \partial_\nu A_\rho
          - \frac{1}{2\alpha}(\partial_\mu A^\mu)^2
          + \bar{\psi}(i \not \! \partial - e \not \! \! A)\psi \ \ , 
\label{lag}
\end{eqnarray}
where $e$ is the gauge coupling constant and $\alpha$ is the gauge-fixing 
parameter.  
The second term in the right-hand side of Eq.(\ref{lag}) is 
the so-called Chern-Simons term.  
It is well-known that the term gives the gauge field the mass $\mu$ without 
breaking the gauge symmetry.  
In fact, a free propagator of the gauge field $iD_{\mu\nu}(p-k)$ derived from 
Eq.(\ref{lag}) is written as 
\begin{equation}
iD_{\mu\nu}(p)=
-i\frac{1}{p^2-\mu^2}\left(g_{\mu\nu}-\frac{p_\mu p_\nu}{p^2}\right)
+\mu\frac{1}{p^2-\mu^2}\frac{1}{p^2}\varepsilon_{\mu\nu\rho}p^{\rho}
-i\alpha\frac{p_\mu p_\nu}{p^4} \ \ ,
\label{D}
\end{equation}
in which we find a massive pole at $p^2=\mu^2$ so that $\mu$ is called 
the topological mass of the gauge field.  

$\psi$ is the two-component fermion field which belongs to the irreducible 
spinor representation in (2+1)-dimensions.  
The Dirac matrices are defined by $\gamma^0=\sigma_3, \gamma^1=i\sigma_1, 
\gamma^2=i\sigma_2$ with diag$(g^{\mu\nu})=(1,-1,-1)$ where $\sigma_i$'s 
(i=1, 2, 3) are the Pauli matrices.  
The $\gamma^\mu$'s satisfy relations as 
$\{ \gamma^\mu, \gamma^\nu \}=2g^{\mu \nu}$, $\gamma^\mu \gamma^\nu = -i 
\epsilon^{\mu \nu \rho} \gamma_\rho + g^{\mu \nu}$ and $tr[\gamma^\mu 
\gamma^\nu ] = 2g^{\mu \nu}$.  
In this representation, there does not exist a matrix which anti-commutes 
with all of $\gamma^\mu$'s so that we cannot define the chiral 
transformation.  
This is a specific aspect of the odd-dimensional space-time.  
In even-dimensions, the chiral symmetry requires that a fermion is 
massless.  
In odd-dimensions, the chiral symmetry itself does not exist.  
Instead, the mass term of the fermion is forbidden by parity symmetry.  
The parity transformation in (2+1)-dimensions is defined as 
$ x=(t, x, y) \rightarrow x'=(t, -x, y), 
\psi(x) \rightarrow \gamma^1 \psi(x'), 
A^0 (x) \rightarrow A^0 (x'), 
A^1 (x) \rightarrow - A^1 (x'), 
A^2 (x) \rightarrow A^2 (x').  $
Under the parity transformation, the mass term of the fermion and the 
Chern-Simons term change their signs.  
Thus the mass terms of both the fermion and the gauge field are forbidden by 
the parity symmetry.  
We study how the breaking of parity by the topological mass affects the 
mass generation of the fermion.  

\section{Fermion self-energy in perturbation}

Before proceeding to a nonperturbative analysis, it would be useful to see a 
fermion self-energy in the lowest order of perturbation in order to make 
prominent why we need a nonperturbative analysis to see the behaviour 
in the $\mu \rightarrow 0$ limit.  
The fermion self-energy in the one-loop approximation, $\Sigma^{(1)}(p)$, is 
expressed as 
\begin{equation}
\Sigma^{(1)}(p)=\int \frac{d^3k}{(2\pi)^3}
(-ie\gamma^\mu) iS_F(k)(-ie\gamma^\nu)
 iD_{\mu\nu}(p-k) \ \ ,
\label{self1}
\end{equation}
where $iS_F(p)$ is a free fermion propagator as 
\begin{equation}
iS_F(p)=\frac{i}{\not{\hspace{-0.5mm}p}} \ \ ,
\label{freefermion}
\end{equation}
and $iD_{\mu\nu}(p-k)$ is a free propagator of the gauge field given 
in Eq. (\ref{D}).  

The allowed form of the fermion propagator in the relativistic theory is 
written as
\begin{equation}
iS^{(1)}_F(p) = \frac{i}{A^{(1)}(p)\not{\hspace{-0.5mm}p}-B^{(1)}(p)} 
              = \frac{i}{\not{\hspace{-0.5mm}p}- i \Sigma^{(1)}(p)} \ \ ,
\label{SF1}
\end{equation}
where $A^{(1)}(p)$ and $B^{(1)}(p)$ are functions of $\sqrt{p_\mu p^\mu}$, 
while $\Sigma^{(1)}(p)$ depends on each $p_\mu$.  
$A^{(1)}(p)^{-1}$ is the wave function renormalization and 
$B^{(1)}(p)/A^{(1)}(p)$ is a mass induced by dynamical effects at the 
momentum scale $p$.  
The so-called dynamical mass $m_{phys}$ is defined by 
$m_{phys}=B^{(1)}(0)/A^{(1)}(0)$ as usual.
It is useful to notice the relations as 
\begin{equation}
tr\left[\Sigma^{(1)}(p)\right]=-2iB^{(1)}(p), \ \ 
tr\left[\not{\hspace{-0.5mm}p}\Sigma^{(1)}(p)\right]
=2i\{A^{(1)}(p)-1\}p^2.
\label{trace}
\end{equation}

We substitute Eqs.(\ref{freefermion}) and (\ref{D}) into 
Eq.(\ref{self1}) and use Eq.(\ref{trace}).  
We change the metric to the Euclidean one by the Wick rotation as 
$(k^0,\vec{k}) \rightarrow (ik^0,\vec{k})$ and $(p^0,\vec{p}) \rightarrow 
(ip^0,\vec{p})$.  
Then $k^2$ and $p^2$ are replaced by $-k^2=-(k^0)^2-(k^1)^2$ and 
$-p^2=-(p^0)^2-(p^1)^2$.  
After that,  we transform the integral variables $k^\mu$'s to the polar 
coordinates $(k, \theta,\phi)$.  
The angular integration on $\theta$ and $\phi$ can be done explicitly.  
Finally we obtain 
\begin{eqnarray}
B^{(1)}(p)&=&\frac{e^2}{8\pi^2 p}\int^\infty_0 dk \frac{1}{k}
\left[ - \frac{1}{\mu}(p^2-k^2)\ln\frac{(p+k)^2}{(p-k)^2} 
\right. 
\nonumber \\
&+&
\left. 
\frac{1}{\mu}(p^2-k^2+\mu^2)\ln\frac{(p+k)^2+\mu^2}{(p-k)^2+\mu^2}
\right] , 
\label{B1} \\
A^{(1)}(p)&=&1+\frac{e^2}{8\pi^2p^3}\int^\infty_0 dk \frac{1}{k} 
\left[ -2pk(\alpha+1) 
\right. 
\nonumber \\
&+& \left.
\left\{\frac{1}{2\mu^2}(p^2-k^2)^2
+ \frac{1}{2}\alpha(p^2+k^2)\right\}\ln\frac{(p+k)^2}{(p-k)^2}
\right.
\nonumber \\
&+& \left.
\left\{\frac{1}{2}\mu^2-\frac{1}{2\mu^2}(p^2-k^2)^2\right\}
\ln\frac{(p+k)^2+\mu^2}{(p-k)^2+\mu^2}
\right] \ \ .
\label{A1}
\end{eqnarray}

The dynamical mass of fermion is defined in the infrared limit so that we 
are interested in the behaviour of $A^{(1)}(p)$ and $B^{(1)}(p)$ in this 
limit.  
In the region of $p \ll 1$, Eqs.(\ref{B1}) and (\ref{A1}) are written as
\begin{eqnarray}
B^{(1)}(p)&=&\frac{e^2}{\pi^2}\int^\infty_0 dk \left[\frac{\mu}{k^2+\mu^2}
+O(p^2) \right]\\
A^{(1)}(p)&=&1+\frac{e^2}{\pi^2}\int^\infty_0 dk
\left[\frac{1}{3}\left\{\frac{1}{k^2}\alpha-2\frac{\mu^2}{(k^2+\mu^2)^2}
\right\}
+O(p^2)\right].
\end{eqnarray}
This infrared approximation makes the integration on k possible and we have 
\begin{eqnarray}
B^{(1)}(0)=\frac{e^2}{2\pi}\frac{|\mu|}{\mu} \ , \ \ 
A^{(1)}(0)=1-\frac{e^2}{6\pi}\frac{|\mu|}{\mu^2}+\frac{e^2}{3\pi^2}
\frac{\alpha}{\epsilon} \ \ ,
\label{Per}
\end{eqnarray}
where $\epsilon$ is the infrared cutoff in the integration on $k$.  
We notice that $A^{(1)}(0)$ is free from the infrared divergence only 
in the Landau gauge.  

At first sight, we can see the non-analytic $\mu$-dependence of $A^{(1)}(0)$ 
and $B^{(1)}(0)$ in Eq. (11):   
$B^{(1)}(0)$ depends on the sign of $\mu$ so that the value of $B^{(1)}(0)$ 
at $\mu=0$ is not well defined.  
$A^{(1)}(0)$ is singular at $\mu=0$.  
Therefore the theory with the Chern-Simons term may not be connected to the 
theory without the Chern-Simons term at least in the lowest order 
perturbation.  
However this consideration is not enough. 

The theory has the two dimensional parameters $e^2$ and $\mu$ which have 
the dimension of mass.  
We take the $p \rightarrow 0$ limit to define the dynamical mass and also the 
$\mu \rightarrow 0$ limit to see the linking of the theories with and without 
the Chern-Simons term.  
Then we have to fix the scale of $e^2$ in taking these limits.  
Thus the correct procedures of taking the limits are given by 
$\frac{|{\vec p}|}{e^2} \rightarrow 0$ and $\frac{\mu}{e^2} \rightarrow 0$.  
Here we notice that the $\mu \rightarrow 0$ limit in fixing $e^2$ is 
equivalent to the strong coupling limit.  
It means that Eq. (11) obtained in the perturbation is not available any more 
and the non-perturbative analysis is needed to investigate the limit.  
This situation motivates us to study the $\mu \rightarrow 0$ limit of the 
dynamical fermion mass by a non-perturbative method.  
This issue is extensively studied in the successive sections.  

\section{Schwinger-Dyson equation}

In this section, we proceed to a nonperturbative analysis, where we use 
the Schwinger-Dyson technique to evaluate the dynamical mass of the fermion.  
The Schwinger-Dyson equation for the fermion self-energy $\Sigma(p)$ is 
written as
\begin{eqnarray}
\Sigma(p)=(-i e)^2 \int \frac{d^3k}{(2\pi)^3} \ \gamma^\mu \ 
  i S'_F (k) \ \Gamma^\nu(k,p-k) \ i D'_{\mu\nu}(p-k) \ \ .
\label{SD}
\end{eqnarray}
$\Gamma^\nu(k,p-k)$ is a full vertex function and $D'_{\mu\nu}(p-k)$ is a 
full propagator of the gauge field.  
$S'_F(p)$ is the full propagator of the fermion field which is  written as  
\begin{eqnarray}
i S'_F(p)=\frac{i}{A(p)\not \hspace{-0.8mm}p - B(p)}
        =\frac{i}{\not \hspace{-0.8mm}p-i\Sigma(p)} \ \ ,
\label{SF}
\end{eqnarray}
which includes the full correction beyond the perturbative fermion propagator 
given in Eq. (\ref{SF1}).  

To analyze Eq.(\ref{SD}) further, we need to introduce any suitable 
approximation.  
In this paper, we limit ourselves to use the lowest ladder approximation 
where the full propagator of the gauge field and the full vertex are replaced 
by the free propagator and the bare vertex respectively as 
\begin{eqnarray}
i D'_{\mu\nu}(p-k) \approx i D_{\mu\nu}(p-k) \ , \ \ 
\Gamma^\nu(k,p-k) \approx \gamma^\nu \ \ ,  
\label{ladder}
\end{eqnarray}
where $i D_{\mu \nu}$ has been given in Eq.(\ref{D}).  
Thus the Schwinger-Dyson equation in the lowest ladder approximation becomes 
\begin{eqnarray}
\Sigma(p)=(-i e)^2\int\frac{d^3k}{(2\pi)^3}
        \gamma^\mu \,i S'_F(k)\gamma^\nu \,i D_{\mu\nu}(p-k) \ \ .
\label{SDld}
\end{eqnarray}

We substitute Eqs.(\ref{D}) and (\ref{SF}) into Eq.(\ref{SDld}).  
Following the same steps used in getting Eqs.(\ref{B1}) and (\ref{A1}), we 
finally obtain the coupled integral equations as 
\begin{eqnarray}
B(p) &=& \frac{e^2}{8\pi^2p} \int_{0}^{\infty}dk 
            \frac{k}{A(k)^2 k^2+B(k)^2} \left[ \left\{\alpha B(k) - 
            \frac{1}{\mu}(p^2-k^2) A(k) \right\} 
            \ln\frac{(p+k)^2}{(p-k)^2} \right. \nonumber \\
        &+& \left. \left\{\frac{1}{\mu}(p^2-k^2) A(k) +\mu A(k) 
         +2 B(k) \right\} \ln\frac{(p+k)^2+\mu^2}{(p-k)^2+\mu^2} \right] 
         \ \ , 
\label{B} \\
A(p) &=& 1+\frac{e^2}{8\pi^2p^3}\int_{0}^{\infty}dk
            \frac{k}{A(k)^2 k^2+ B(k)^2} 
            \left[ -2pk(\alpha+1) A(k) 
            \right. 
            \nonumber \\
        &+& \left. \left\{\frac{1}{2\mu^2}(p^2-k^2)^2 A(k) 
          + \frac{1}{\mu}(p^2-k^2) B(k) 
          + \frac{1}{2}\alpha(p^2+k^2) A(k) \right\} 
            \ln\frac{(p+k)^2}{(p-k)^2}  
            \right. 
            \nonumber \\
        &+& \left. \left\{\frac{1}{2}\mu^2 A(k) 
          - \frac{1}{2\mu^2}(p^2-k^2)^2 A(k)
          + \mu B(k) - \frac{1}{\mu}(p^2-k^2) B(k) \right\} 
        \right. 
        \nonumber \\
        &\times&
        \left.
            \ln\frac{(p+k)^2+\mu^2}{(p-k)^2+\mu^2} \right]
            ,
\label{A}
\end{eqnarray}
which contain only the integration on the radial variable $k$.  
We solve these equations by an approximated analytical method in this section 
and also numerically by using an iteration method in the following section. 

The limitation to the lowest ladder approximation is really unsatisfactory.  
This is mainly because the Schwinger-Dyson equation in the theory with the 
Chern-Simons term is complicated than the one in the theory without the 
Chern-Simons term, which will be explained soon later.  
In the situation, we think that it would be better to accumulate our 
experience in analysing the theory under a rather simpler approximation and to 
make a confidential first step as a springboard which is useful for a more 
extended next analysis.  
Even in the lowest ladder approximation, we can get many important 
informations. 
The results so found within the approximation might be denied in a more 
sophisticated approximation.  
But the analysis in the lowest ladder approximation will tell us what we 
should be careful for.  
Of course, there is no guarantee that the lowest ladder approximation is 
enough for our purpose.  
We will proceed our analysis beyond the lowest ladder approximation in future. 

We can check easily that Eqs.(\ref{B}) and (\ref{A}) reduce to the 
Schwinger-Dyson equations in $QED_3$ without the Chern-Simons term if we 
put the topological mass $\mu$ equal to zero.  
In fact, taking the limit as $\mu \rightarrow 0$ in Eqs. (\ref{B}) and 
(\ref{A}), we obtain 
\begin{eqnarray}
B(p)&=&(\alpha+2) \frac{e^2}{8 \pi^2 p} \int^\infty_0 dk \frac{kB(k)} 
     {A(k)^2 k^2 + B(k)^2} \ln\frac{(p+k)^2}{(p-k)^2} \ \ , 
\label{BwoCS} \\
A(p)&=&1 - \alpha \frac{e^2}{4 \pi^2 p^3} \int^\infty_0 dk \frac{kA(k)} 
     {A(k)^2 k^2 + B(k)^2} \left[ pk - \frac{p^2+k^2}{4} 
      \ln\frac{(p+k)^2}{(p-k)^2} \right] \ \ ,
\label{AwoCS}
\end{eqnarray}
which is the Schwinger-Dyson equations in the lowest ladder approximation
derived in $QED_3$ without Chern-Simons term.  
We can see that there exists the specific gauge where the wave function 
renormalization is absent.  
Thus in the Landau gauge$(\alpha=0)$, Eq.(\ref{AwoCS}) gives us the simple 
solution as $A(p)=1$ and the problem reduces to solve Eq. (\ref{BwoCS}) 
with $A(p)=1$.  

In the case with the Chern-Simons term, as is seen in Eqs.(\ref{B}) and 
(\ref{A}), there does not exist such a specific gauge where the wave function 
is not renormalized.  
So far we cannot find a self-evident reason that the Landau is still specific 
in $QED_3$ with the Chern-Simons term, it must be fair to study Eqs. (\ref{B}) 
and (\ref{A}) for various values of the gauge-fixing parameter $\alpha$.  

While the Schwinger-Dyson equations (\ref{B}) and (\ref{A}) in the 
Maxwell-Chern-Simons $QED_3$ reduce to the equations (\ref{BwoCS}) and 
(\ref{AwoCS}) in $QED_3$ without the Chern-Simons term under the "naive" 
$\mu \rightarrow 0$ limit before integration, it is not so obvious whether 
the solutions of Eqs. (\ref{B}) and (\ref{A}) tend to the ones of Eq. 
(\ref{BwoCS}) and (\ref{AwoCS}).  
We cannot exclude a possibility that the interchange of the $\mu \rightarrow 
0$ limit and the integration is not allowed because of a nontrivial nature of the 
integration kernels as distribution functions.  
In addition, we know that the Schwinger-Dyson equations (\ref{BwoCS}) and 
(\ref{AwoCS}) have two solutions.  
One is trivial ($B(p)=0$) and the other is non-trivial.  
It is interesting to find how the solution of Eqs. (\ref{B}) and (\ref{A}) behaves 
in the $\mu \rightarrow 0$ limit.  

It is very useful if we can estimate $A(0)$ and $B(0)$ analytically even under 
a fairly crude approximation.  
The kernels of these integral equations are dumped rapidly as the integral 
variable $k$ increases so that the contribution from $k \approx 0$ is the 
most dominant one in the integrals.  
We approximate $A(k)$ and $B(k)$ by $A(0)$ and $B(0)$ in the integrals.  
We call this approximation ''the $constant$ approximation''.  
Of course this approximation might be too crude for our purpose and we only 
use the result as reference in the numerical analysis.  
Under this approximation, we can perform the remaining radial integration 
and obtain
\begin{eqnarray}
B(0)=\frac{e^2}{2\pi} + \frac{e^2}{12\pi}\alpha \ , \ \ 
A(0)=1+\frac{2\alpha}{\alpha+6} \ \ , 
\label{BAcnst}
\end{eqnarray}
where we have considered the case of $\mu>0$.  

From Eq.(\ref{BAcnst}), we can see that the dependence 
of $B(0)$ and $A(0)$ on the gauge-fixing parameter, the coupling constant 
and the topological mass has the following peculiar features:
$B(0)$ depends linearly on $\alpha$.  
In the Landau gauge ($\alpha=0$), $A(0)=1$ and $B(0)=e^2/2\pi$.  
$A(0)=1$ is favourable for us because $A(p)=1$ means that the 
Ward-Takahashi identity is satisfied.  
$A(0)$ does not depend on $e^2$.  
It means that the deviation of $A(0)$ from 1 is independent of the coupling 
constant.  
This is crucially different from the perturbative result given by 
Eq.(\ref{Per}) where the deviation is proportional to $e^2$.  
On the other hand, $B(0)$ is proportional to $e^2$.  
We recognize that  there is no dependence on the topological mass 
$\mu$ in Eq.(\ref{BAcnst}).  
In fact, if we apply the constant approximation to the case without the 
Chern-Simons term, we obtain the same results as Eq.(\ref{BAcnst}).
It means that the amount of the explicit parity breaking in the gauge 
sector by the topological mass does not affect the dynamical mass in 
the fermion sector in the constant approximation.  

Now we proceed to a more precise numerical evaluation in the next section.  

\section{Numerical analysis}

\subsection{Nontrivial solutions and gauge dependence}

We solve the two coupled integral equations (\ref{B}) and (\ref{A}) 
numerically by using a method of iteration.  
First we substitute trial functions into $A(k)$ and $B(k)$ on the right-hand 
sides of Eqs. (\ref{B}) and (\ref{A}) 
and then calculate the integrals numerically.  
The outputs so obtained, $A(p)$ and $B(p)$, are substituted back to the 
right-hand sides until the outputs coincide with the inputs.  
Finally we obtain convergent functions $A(p)$ and $B(p)$, which satisfy the 
integral equations, if there exist any solutions in Eqs. (\ref{B}) and 
(\ref{A}).

We have obtained the nontrivial solutions for the various values of the gauge 
parameter $\alpha$.  
We have found that $A(p)$ is almost a constant and its value is fairy close 
to 1 in the Landau gauge ($\alpha=0$) even for a very small $\mu$.  
Within the ladder approximation, no vertex correction is taken into account 
so that the wave function also should not be renormalized and $A(p)$ should be 
equal to 1.  
This means that the Ward-Takahashi identity is satisfied and the approximation 
is consistent with the gauge invariance.  
While we do not show the details here since the situation on this point is same 
as the one in ref. \cite{MNU}, the Landau gauge is still the best gauge.  
Therefore we present the results obtained in the Landau gauge hereafter.

\subsection{Dependence on the topological mass}

We are most interested in the dependence of the dynamical fermion 
mass on the topological mass of the gauge field.  
In the constant approximation, both $A(0)$ and $B(0)$ do not depend on the 
topological mass as seen in Eq. (\ref{BAcnst}).    
We estimate them by the more precise numerical method.  

We have studied the dependence of $A(0)$ on the dimensionless parameter 
$\hat{\mu}$ which is defined by $\hat{\mu}=\mu / e^2 $.  
($A(0)$ has no dimension.  $\mu$ and $e^2$ have the dimension of mass.)  
We have found that the deviation of $A(0)$ from $1$ is less than 1 \%.  
We may say that $A(0)$ is almost equal to 1 even in the extended region of 
$\hat{\mu}$ which is wider than the one in Ref. \cite{MNU}.  

The $\hat{\mu}$-dependence of $B(0)$ is nontrivial.  
One of our motivations is to see whether or not the Maxwell-Chern-Simons 
$QED_3$ is smoothly connected to $QED_3$ without the Chern-Simons term 
in the $\hat{\mu} \rightarrow 0$ limit.  
Our numerical calculations show that very small meshes are needed 
to obtain reliable values of $B(0)$ in the region $\hat{\mu} \ll 1$.  
Because of the limitation of our machine ability, we take  another strategy 
different from our previous work ~\cite{MNU}.  
We estimate $B(0)$ on meshes of zero-width by an extrapolation from some 
$B(0)$'s on meshes of different finite-widths.  

\begin{figure}[h]
\epsfysize=7cm
\centerline{\epsfbox{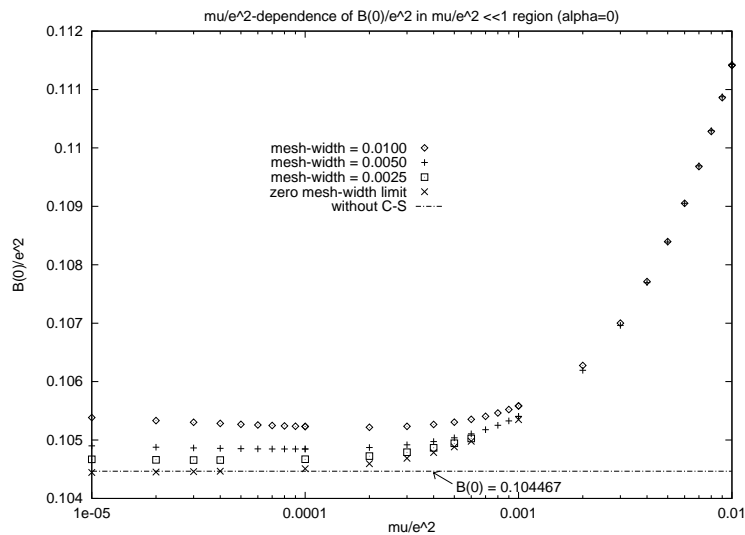}}
\caption
{The precise check of the $\hat{\mu}$-dependence of $B(0)/e^2$ in the region of 
$10^{-5} \leq \hat{\mu} \leq 10^{-2}$.
}
\end{figure}

In Fig.1, we show the $\hat{\mu}$-dependence of $B(0)$ in the region of 
$10^{-5} \le \hat{\mu} \le 10^{-2}$.  
($B(0)$ has the dimension of mass so that we plot $B(0)/e^2$ with no mass 
dimension.)  
$B(0)$'s on meshes of finite-width show an abnormal behaviour in the region 
$\hat{\mu} \ll 1$ that $B(0)$'s depart from $B(0)$ of $QED_3$ without the 
Chern-Simons term as $\hat{\mu}$ decreases.  
This behaviour is improved by the extrapolation.  
We have estimated $B(0)$ on meshes of zero-width by the curve fitting of three 
$B(0)$'s on meshes of finite-widths.  
$B(0)$'s obtained by the extrapolation smoothly tend to the value of $B(0)$ 
in $QED_3$ without the Chern-Simons term.  

The whole shape of $B(0)$ in the region of $10^{-5} \leq \hat{\mu} \leq 10^4$, 
combining with the result in Ref. \cite{MNU}, is given in Fig.2.  
$B(0)$ is almost a constant in the region of $\hat{\mu} > 10$ and decreases 
rapidly in the region $\hat{\mu}= 1.0 \sim 0.01$.  
In the region of $\hat{\mu} < 0.001$,  $B(0)$ becomes almost a constant again.  
The upper dotted line in Fig.2 is the result obtained in the constant 
approximation (Eq. (\ref{BAcnst})) and also in the lowest order perturbation 
(Eq. (\ref{Per})).  
The lower dotted line shows the value obtained by a nonperturbative 
calculation in the case without Chern-Simons term.~\cite{HMH}  
In other words, B(0) reproduces the result of perturbation in the region of 
$e^2 \ll \mu$ while it is close to the nonperturbative result of $QED_3$ 
without the Chern-Simons term in the region of $e^2 \gg \mu$.  

\begin{figure}[h]
\epsfysize=7cm
\centerline{\epsfbox{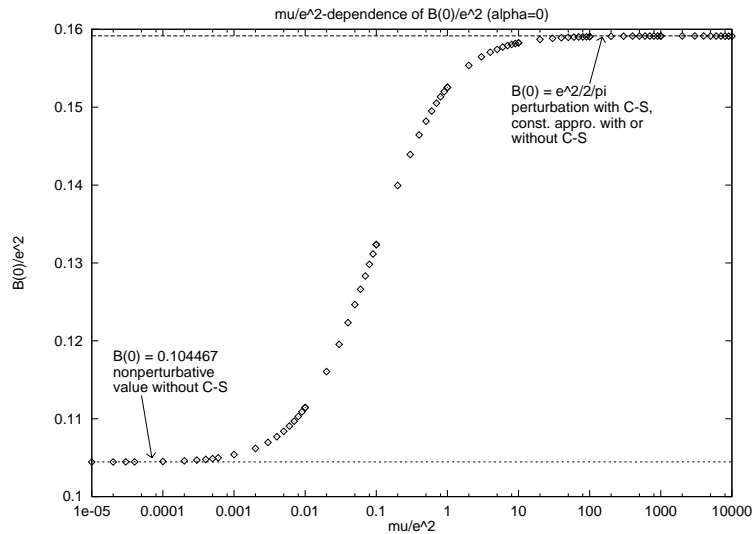}}
\caption
{The $\hat{\mu}$-dependence of $B(0)/e^2$ in the region of $10^{-5} \leq \hat{\mu} 
\leq 10^4$.
}
\end{figure}

\subsection{Parity condensate}

Another important quantity of indicating to what extent the parity symmetry 
is broken by the topological mass is a parity condensate, which is a 
gauge invariant order parameter of a vacuum.  
We evaluate the parity condensate as a function of 
the topological mass in order to know the nature of the vacuum.  
The parity condensate is defined by 
\begin{eqnarray}
<\bar{\psi}\psi> = \lim_{x\rightarrow 0} tr [ iS'_F(x) ] \ \ ,
\label{CDdef}
\end{eqnarray}
where $iS'_F(x)$ is a propagator in real space-time coordinates, which is 
related to Eq.(\ref{SF}) by the Fourier transformation as 
\begin{eqnarray}
i S'_F(x) = \int \frac{d^3p}{(2\pi)^3} i S'_F(p) {\rm e}^{-ipx} \ \ .
\label{FRR}
\end{eqnarray}
By combining Eqs.(\ref{SF}), (\ref{CDdef}) and (\ref{FRR}) and using the Wick 
rotation, we obtain
\begin{eqnarray}
<\bar{\psi} \psi>= \frac{1}{\pi^2} \int^\infty_0 dk 
                   \frac{k^2 B(k)}{A(k)^2 k^2 + B(k)^2} \ \ .
\label{CD}
\end{eqnarray}
We have known the numerical data of $A(p)$ and $B(p)$ for various values of 
$\hat{\mu}$, which has been obtained by solving Eqs. (\ref{B}) and (\ref{A}), 
so that the $\hat{\mu}$-dependence of $<\bar{\psi}\psi>$ is calculable by a 
numerical integral.  

\begin{figure}[h]
\epsfysize=7cm
\centerline{\epsfbox{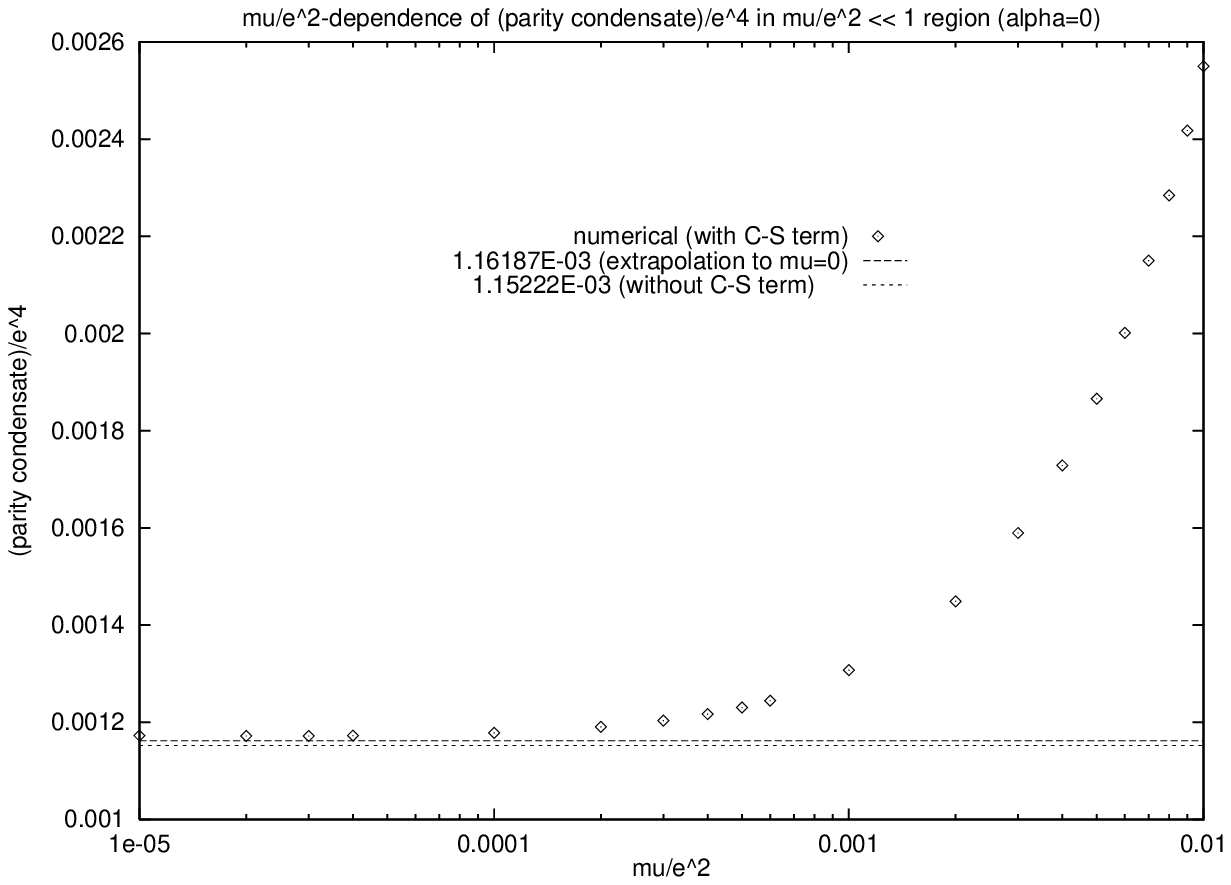}}
\caption
{The $\hat{\mu}$-dependence of $<\bar{\psi}\psi>/e^4$ in the region of $10^{-5} 
\leq \hat{\mu} \leq 10^{-2}$.
}
\end{figure}

In Fig. 3, we show the $\hat{\mu}$-dependence of $<\bar{\psi}\psi>/e^4$ in the 
region of $10^{-5}<\hat{\mu}<10^{-2}$.  
($<\bar{\psi}\psi>$ has the dimension of (mass)$^2$.  
Therefore we plot a dimensionless quantity $<\bar{\psi}\psi>/e^4$.)  
To inspect the $\hat{\mu} \rightarrow 0$ limit, we extrapolate the $B(0)$'s 
in that region to $\hat{\mu}=0$ by a least mean square method.  
The value obtained by the extrapolation is $<\bar{\psi}\psi>/e^4=
1.16187\times 10^{-3}$.  
On the other hand, we also evaluate $B(p)$ in $QED_3$ without the Chern-Simons 
term by solving Eqs. (\ref{BwoCS}) in the Landau gauge ($A(p)=1$).  
Then we calculate the condensate of Eq. (\ref{CD}) numerically and obtain 
$<\bar{\psi}\psi>/e^4=1.15222\times 10^{-3}$.  
Both values coincide within an error less than 1 \%.  
Therefore we may say that the behaviour of $<\bar{\psi}\psi>$ also supports 
that $QED_3$ with the Chern-Simons term reduces to $QED_3$ without the 
Chern-Simons term smoothly in the $\hat{\mu} \rightarrow 0$ limit.  

The $\hat{\mu}$-dependence of $<\bar{\psi}\psi>/e^4$ in the region of $10^{-5} 
\le \hat{\mu} \leq 10^4$ is shown in Fig. 4.  
In the region of $\hat{\mu} \ll 1$, the condensate is almost a constant.  
Around $\hat{\mu} \approx 0.001$, it starts to increase.  
For $\hat{\mu} \gg 1$, the increasing of the condensate is almost linear.  
The parity condensate increases more and more as $\hat{\mu}$ does.  
Notice that there is no saturation for the increasing of $<\bar{\psi}\psi>/e^4$.  

\begin{figure}[h]
\epsfysize=7cm
\centerline{\epsfbox{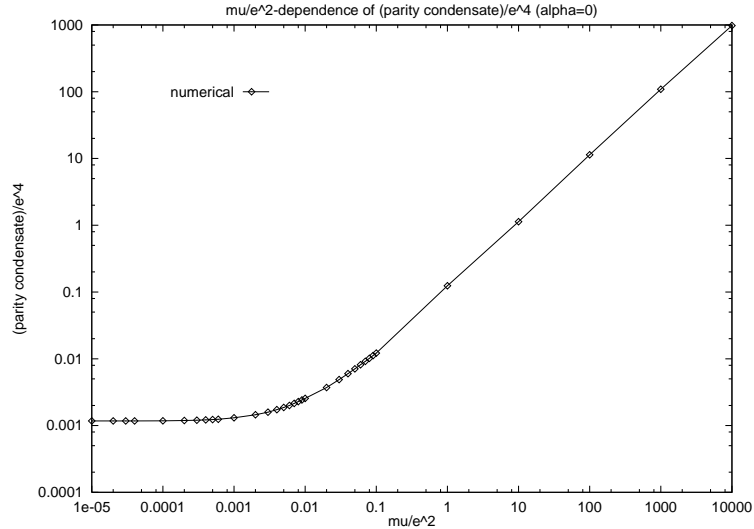}}
\caption
{The $\hat{\mu}$-dependence of $< \bar{\psi} \psi >/e^4$ in the region of $10^{-5} 
\leq \hat{\mu} \leq 10^4$.
}
\end{figure}

\section{Conclusion}

We have studied the dependence of the dynamical fermion mass and the parity 
condensate on the topological mass in the Maxwell-Chern-Simons $QED_3$ 
nonperturbatively by using the Schwinger-Dyson method.  

When the topological mass is larger than the square of coupling constant, 
the value of the topological mass remains to be the one obtained by the 
perturbation.  
As the topological mass decreases, the value is changed to a nonperturbative 
value rapidly.  
The transition from the perturbative value to the nonperturbative one is sharp 
but not critical.  
Though it does not seem to be a phase transition, the inclusion of the 
Chern-Simons changes the nature of the theory drastically.  

Motivated by the behaviour of the fermion self-energy in the 
perturbation, we have studied whether or not $QED_3$ with the Chern-Simons 
term reduces to $QED_3$ without the Chern-Simons term in the zero limit of the 
topological mass.  
We have checked the behaviour of the dynamical mass and the parity condensate 
for the extremely small topological mass in detail.  
The result shows that both quantities tend to the ones of $QED_3$ without 
the Chern-Simons term.  
Thus we conclude that $QED_3$ with the Chern-Simons term reduces to $QED_3$ 
without the Chern-Simons term smoothly in the nonperturbative level.  

In general, an addition of a topological term to an action can give us a 
highly nontrivial deformation of a theory.  
Then it is not self-evident how the theories with or without the topological 
term are connected each other.  
The present work suggests that the linking of the theories should be 
considered in a nonperturbative level.  

\section*{Acknowledgment}
One of the authors (T. M.) would like to thank Professor M. Kenmoku for his 
hospitality at the Nara Women's University.

\end{document}